\documentclass[twocolumn,showpacs,preprintnumbers,amsmath,amssymb]{revtex4}
\usepackage{graphicx}
\begin{document}
\title{Impact of Kondo correlations and spin-orbit coupling on spin-polarized transport in carbon nanotube quantum dot}
\author{D. Krychowski, S. Lipi\'{n}ski}
\affiliation{%
Institute of Molecular Physics, Polish Academy of Sciences\\M. Smoluchowskiego 17,
60-179 Pozna\'{n}, Poland
}%
\date{\today}
\begin{abstract}
Spin polarized transport through a  quantum dot coupled to ferromagnetic electrodes with noncollinear magnetizations  is discussed in terms of nonequilibrium Green functions formalism in the finite-U slave boson mean field approximation. The difference of orientations of the magnetizations of electrodes opens off-diagonal spin-orbital transmission and apart from spin currents of longitudinal polarization also spin-flip currents appear.  We also study equilibrium pure spin current at zero bias and discuss its dependence on magnetization orientation, spin–orbit coupling strength and gate voltage.  Impact of these factors on  tunneling magnetoresistance (TMR) is also undertaken. In general  spin-orbit coupling  weakens TMR, but it can change its sign.\end{abstract}

\maketitle

\section{Introduction}
Spin-dependent coherent electronic transport attracts great interest due to its potential applications in reprogrammable logic devices and quantum computers \cite{Sarma, DiVincenzo, Samarth}.  Carbon nanotubes (CNT) are very promising materials for spintronic applications due to their long spin lifetimes and because they can be contacted with ferromagnetic materials \cite{Kontos}. Additional to spin, orbital degrees of freedom corresponding to clockwise and counterclockwise symmetry of wrapping modes in CNTs \cite{Park} open a new path for quantum manipulation. Due to the  intrinsic spin-orbit interaction (SO), enhanced by curvature,  spin and orbital degrees of freedom are not independent \cite{McEuen}. Our main focus in this article is study of transport in strongly correlated regime. The energy of  SO coupling in CNT quantum dot (CNTQD) is comparable to Kondo energy scale and therefore taking this perturbation into account is important when analyzing many-body effects. For spintronic applications of fundamental importance is tunnel magneto-resistance (TMR),  the relative resistance change with the change of the orientation of polarized electrodes. Recently, there has been an increasing interest of generation of pure spin current (SC) without an accompanying charge current.  The attractive attribute of spin current is that it is associated with a flow of angular momentum, which is a vector quantity. This feature increases  information  capacity transferred through the system.  In contrast to charge flow, spin transport  is almost dissipativeless. The discussion of the dependence of TMR, spin currents and spin accumulation in CNTQD on the strength of magnetic polarization, spin-orbit amplitude, gate voltage and the angle between magnetic moments of the electrodes is the subject of the present publication.

We discuss transport through carbon nanotube quantum dot coupled to the noncollinear spin polarized electrodes (Fig. 1a). CNTQD exhibits fourfold shell structure in the low energy spectrum. For short nanotubes with well separated energy levels it is enough to restrict at low temperatures to the single shell. The system is modelled by two-orbital Anderson model with equal intraorbital and interorbital interaction parameters U and the strength of spin-orbit coupling $\Lambda_{so}$:
\begin{eqnarray}
&&\nonumber{\cal{H}}=\sum_{ls}E_{ls}n_{ls}+\sum_{k\alpha ls}E_{k\alpha s}n_{k\alpha ls}+\sum_{k\alpha ls}(tc^{\dagger}_{kLls}d_{ls}+\\&& t_{ss}(\varphi)c^{\dagger}_{kRls}d_{ls}+t_{s\overline{s}}(\varphi)c^{\dagger}_{kR ls}d_{l\overline{s}}+h.c.)+\\&&\nonumber U\sum_{l}n_{l\uparrow}n_{l\downarrow}+U\sum_{ss'}n_{1s}n_{-1s'},
\end{eqnarray}
where $l=\pm1$, $s=\pm1$ are the orbital and spin indexes. It is assumed that magnetization of the left electrode is oriented along the nanotube axis, whereas magnetization of the right electrode is tilted by angle $\varphi$ (Fig.1a). $c^{\dagger}_{k\alpha ls}$ is the creation operator in the left or right electrode $\alpha = L(R)$, $t$ is hopping integral, $n_{ls}= d^{\dagger}_{ls}d_{ls}$ ($n_{k\alpha ls}=c^{\dagger}_{k\alpha ls}c_{k\alpha ls}$) is  the occupation number operator in CNTQD (lead), $E_{k\alpha ls}$ ($E_{ls}=E_{d}(V_{g})+ls\Lambda_{so}$) denote energies of electrons in the lead (dot). For CNT with the wide bandgap, total spin-orbit contribution to energy (Zeeman-like and orbital-like) can be parametrized by a single effective SO parameter ($\Lambda_{so}$) \cite{McEuen}. The tunneling elemets of the right electrode are given by $t_{ss}=t\cos(\frac{\varphi}{2})$, $t_{1-1}=t\sin(\frac{\varphi}{2})$ and $t_{-11}=-t\sin(\frac{\varphi}{2})$.
To analyze strongly correlation effects in the system we use finite $U$ slave boson mean field approach (SBMFA) of Kotliar and Ruckenstein \cite{Kotliar}. Slave bosons $e$, $p_{ls}$, $d_{20(02)}$, $d_{ss'}$, $t_{ls}$ and $f$ are introduced to describe empty, single , doubly, triple and fully-four occupied states respectively.  Six $d$ bosons project onto states with double occupation of a given orbital $|2,0\rangle$, $|0,2\rangle$ or  onto the  states with single occupation on each orbital $|s,s’\rangle$.
\begin{figure}
\includegraphics[width=0.88\linewidth]{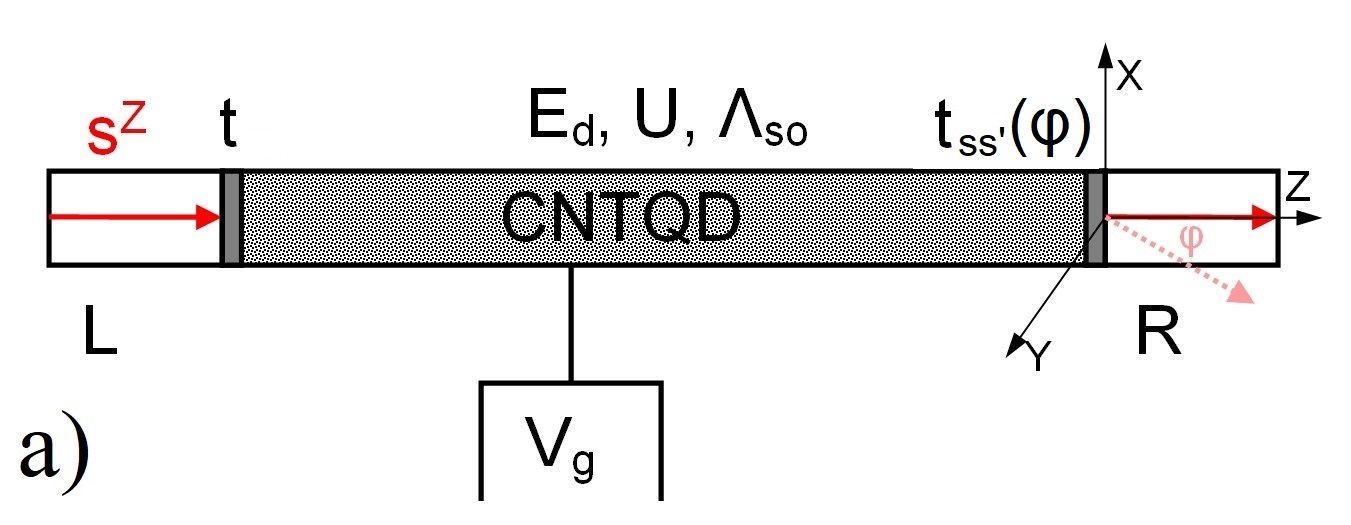}\\
\includegraphics[width=0.44\linewidth]{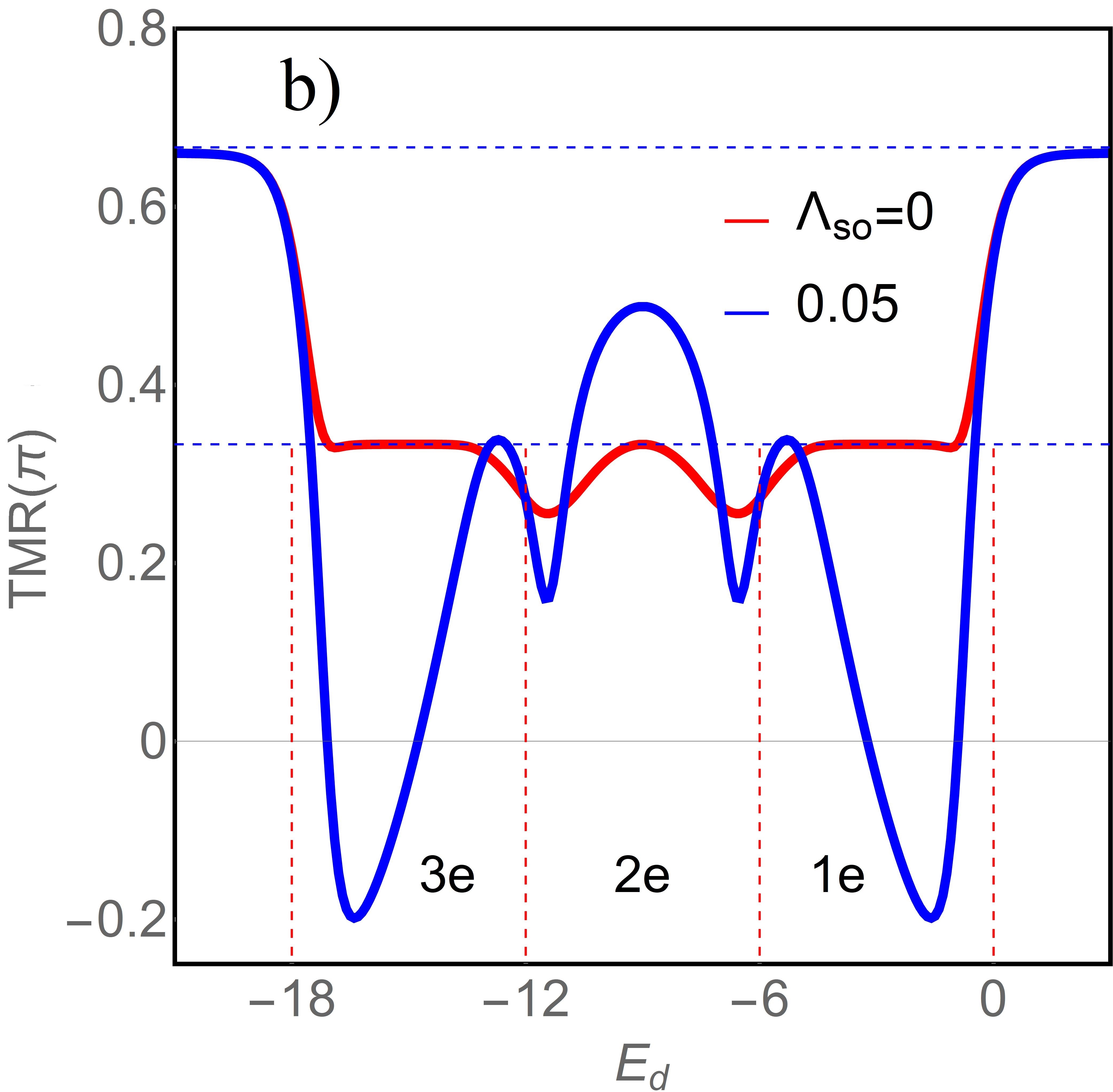}
\includegraphics[width=0.44\linewidth]{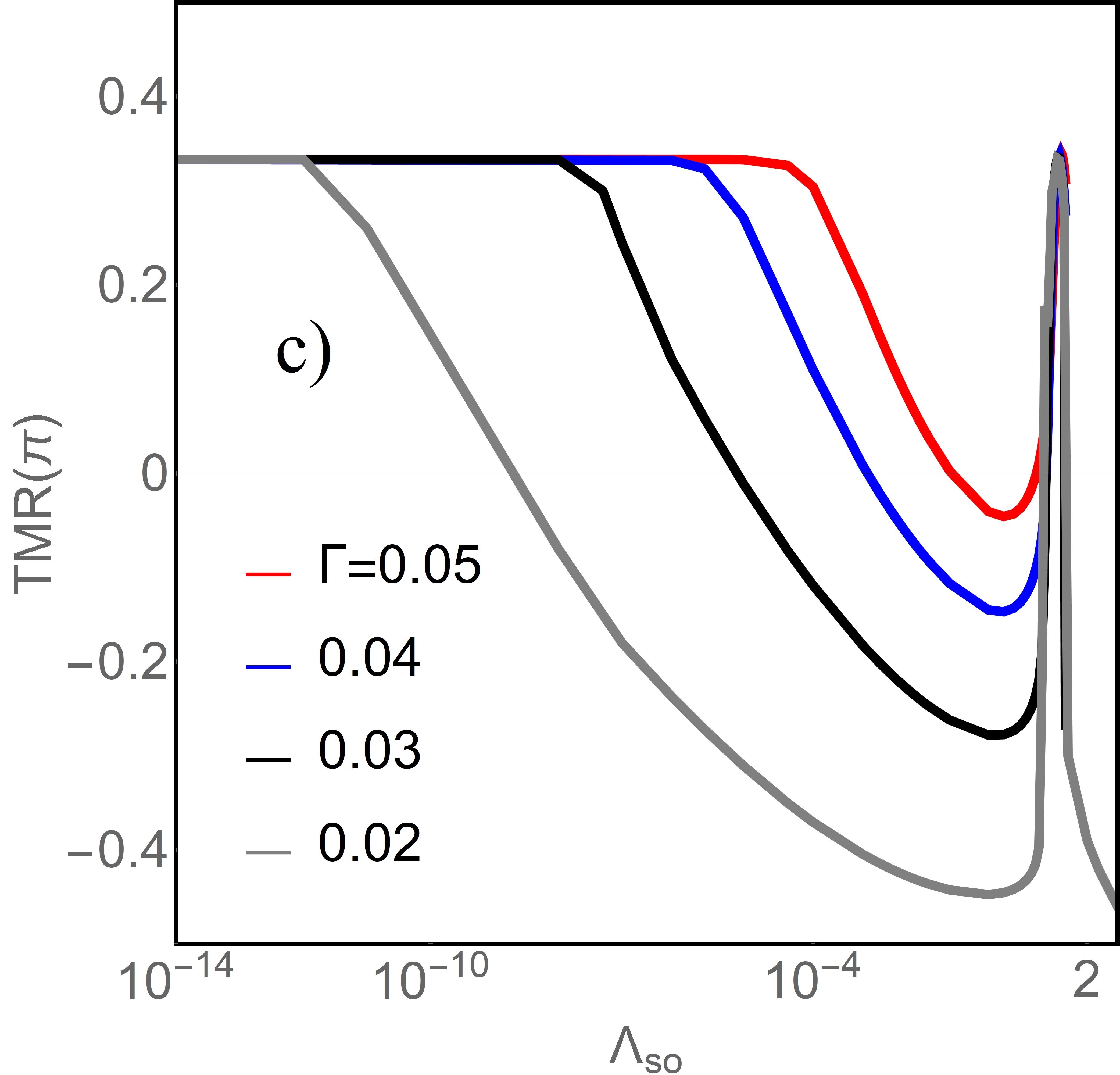}\\
\includegraphics[width=0.44\linewidth]{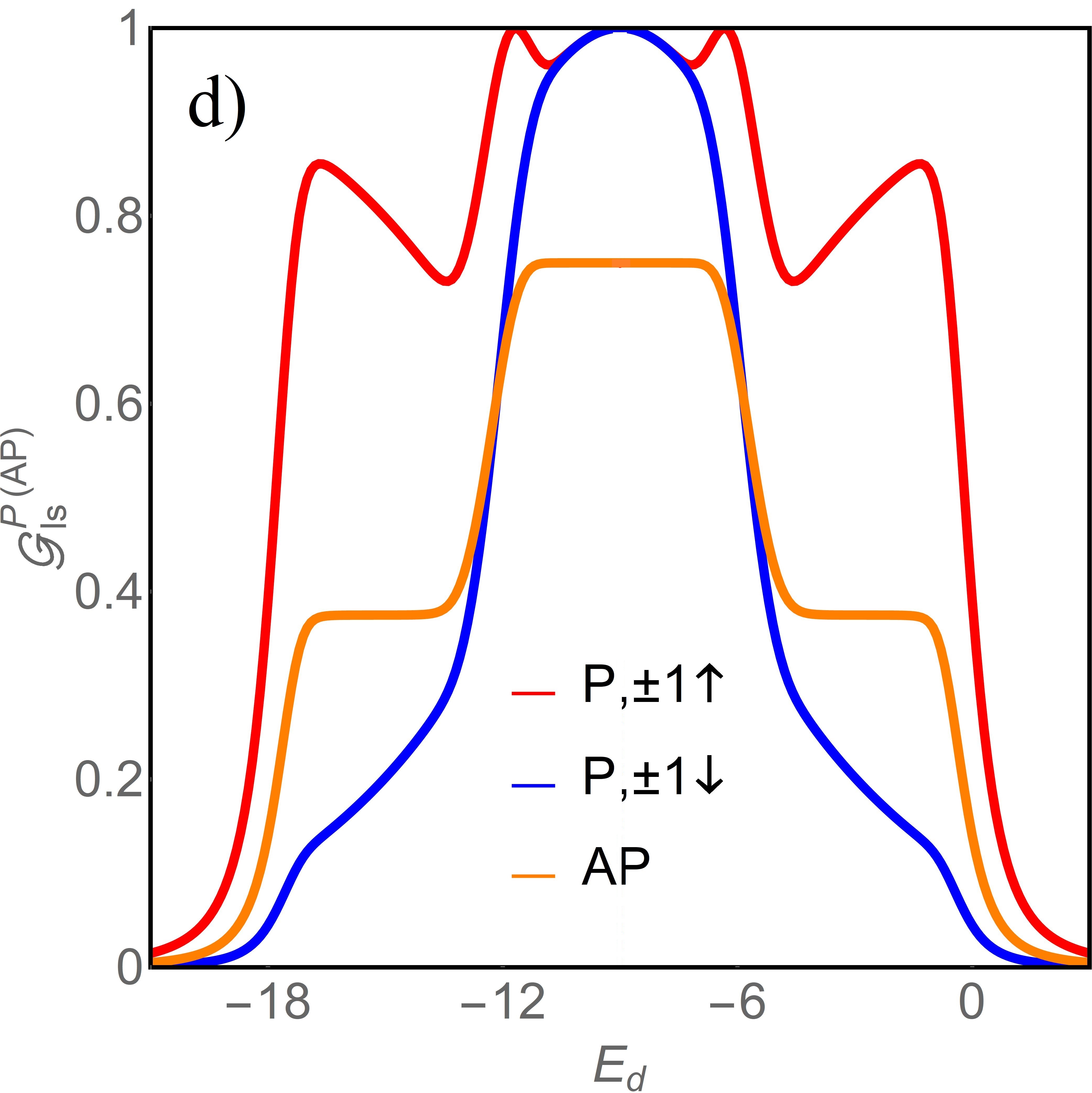}
\includegraphics[width=0.44\linewidth]{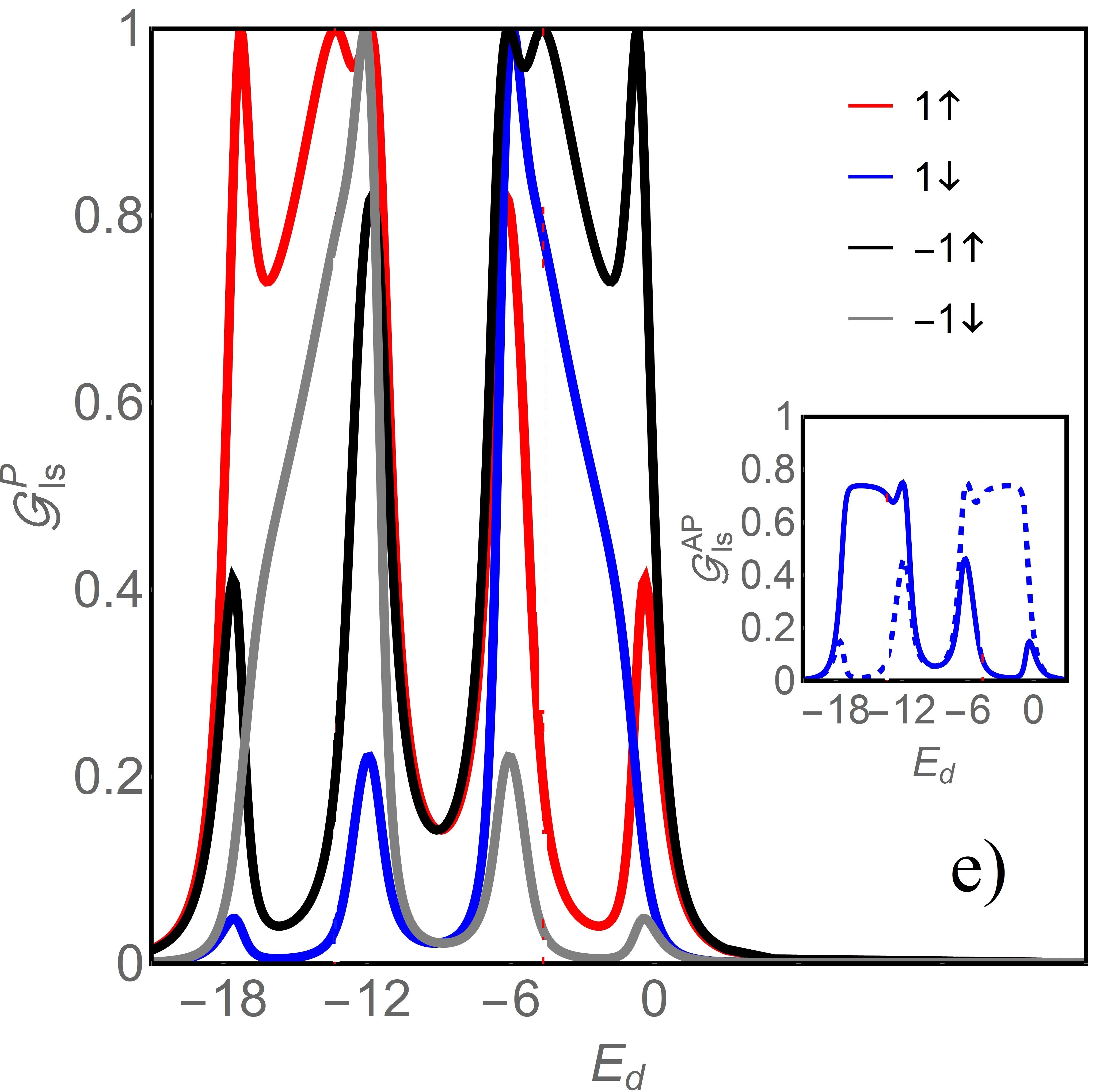}
\caption{\label{fig1} (Color online) (a) – Schematic view of CNTQD with non-colinear spin polarized electrodes, (b) TMR as a function of gate voltage (c) Evolution of TMR with the increase of SO coupling (SOC). (d,e) - Gate-voltage dependencies of spin-orbital resolved conductances in parallel and antiparallel configuration of magnetization in the electrodes without and with SOC in CNTQD $p = 1/2$, (d) $\Lambda_{so} = 0$, (e) $\Lambda_{so} = 0.05$.  Inset of Fig. 1e presents conductances for $1\uparrow,-1\downarrow$ (solid blue line) and $1\downarrow,-1\uparrow$ (dashed blue line) in AP conifiguration.}
\end{figure}
The effective slave boson Hamiltonian (2) reads:
\begin{eqnarray}
&&\nonumber{\cal{\widetilde{H}}}=\sum_{ls}E_{ls}N_{ls}+\sum_{k\alpha ls}E_{k\alpha s}n_{k\alpha ls}+\sum_{k\alpha ls}(tc^{\dagger}_{kLls}z_{ls}f_{ls}+\\&&\nonumber t_{ss}(\varphi)c^{\dagger}_{kRls}z_{ls}f_{ls}+t_{s\overline{s}}(\varphi)c^{\dagger}_{kR ls}z_{l\overline{s}}f_{l\overline{s}}+h.c.)+\\&& U\sum_{lss'}(d^{\dagger}_{l}d_{l}+d^{\dagger}_{ss'}d_{ss'})+3U\sum_{ls}t^{\dagger}_{ls}t_{ls}+6Uf^{\dagger}f
+\\&&\nonumber\sum_{ls}\lambda_{ls}(N_{ls}-Q_{ls})+\lambda(I-1),
\end{eqnarray}
\begin{figure}
\includegraphics[width=0.48\linewidth]{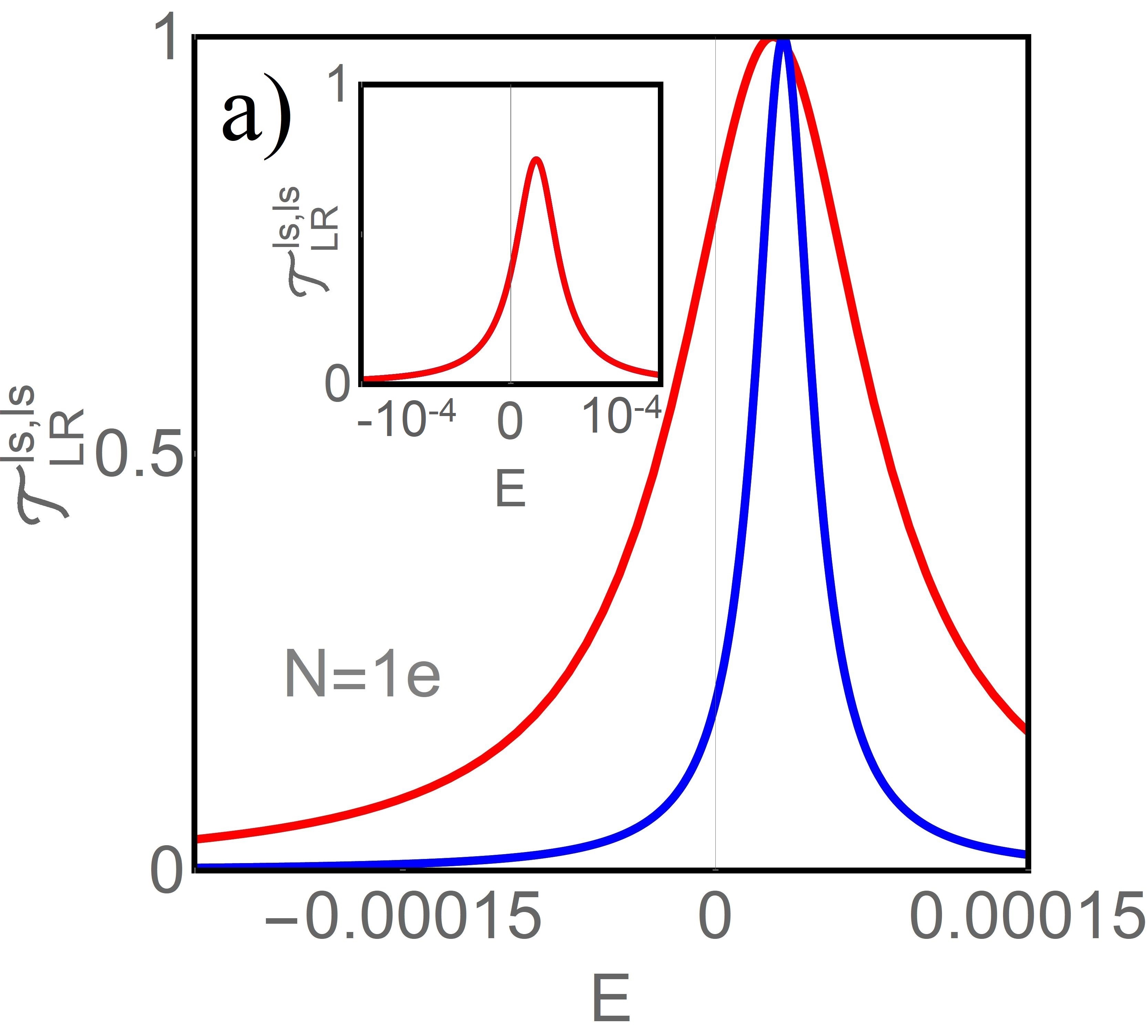}
\includegraphics[width=0.48\linewidth]{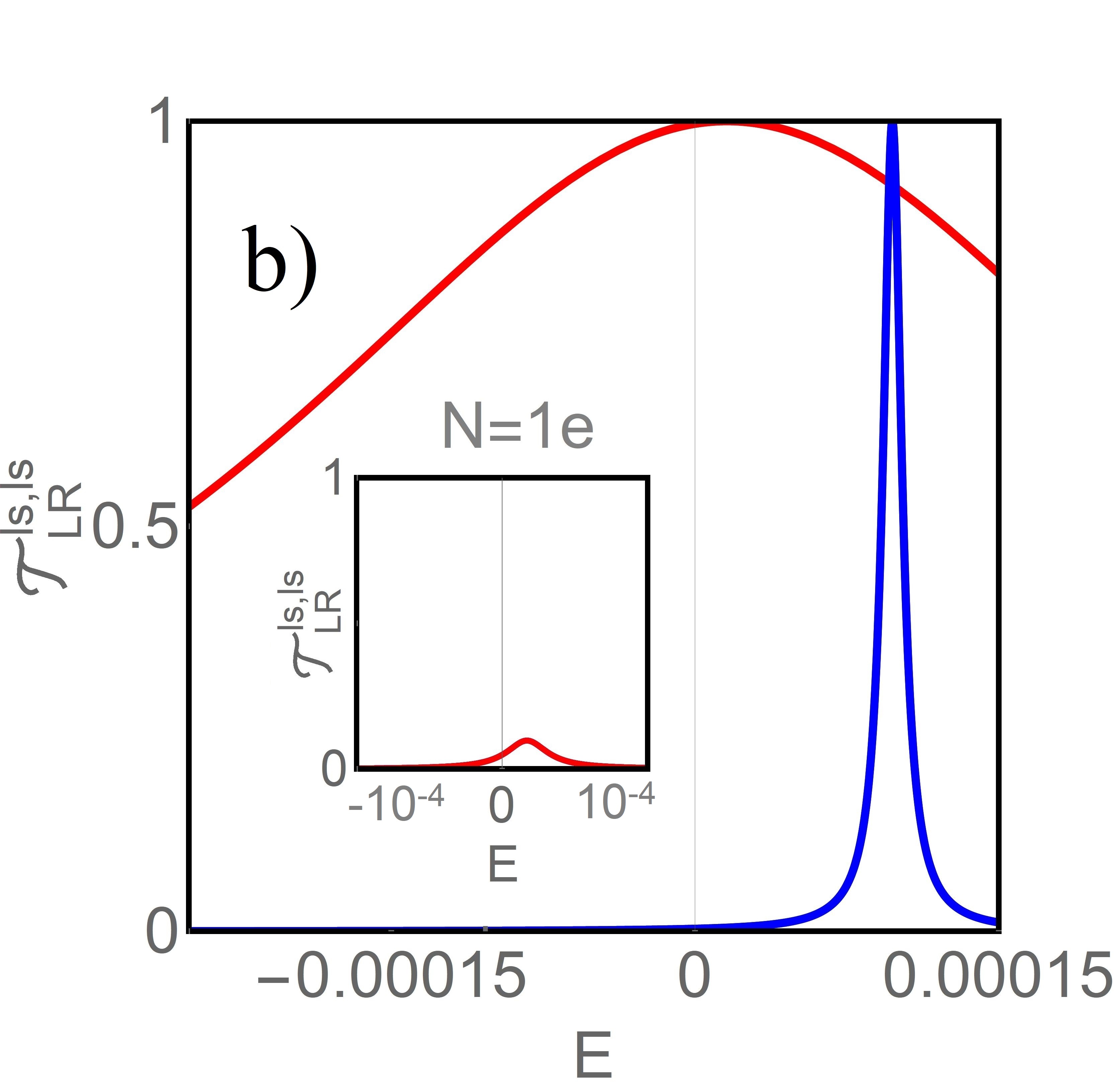}\\
\includegraphics[width=0.48\linewidth]{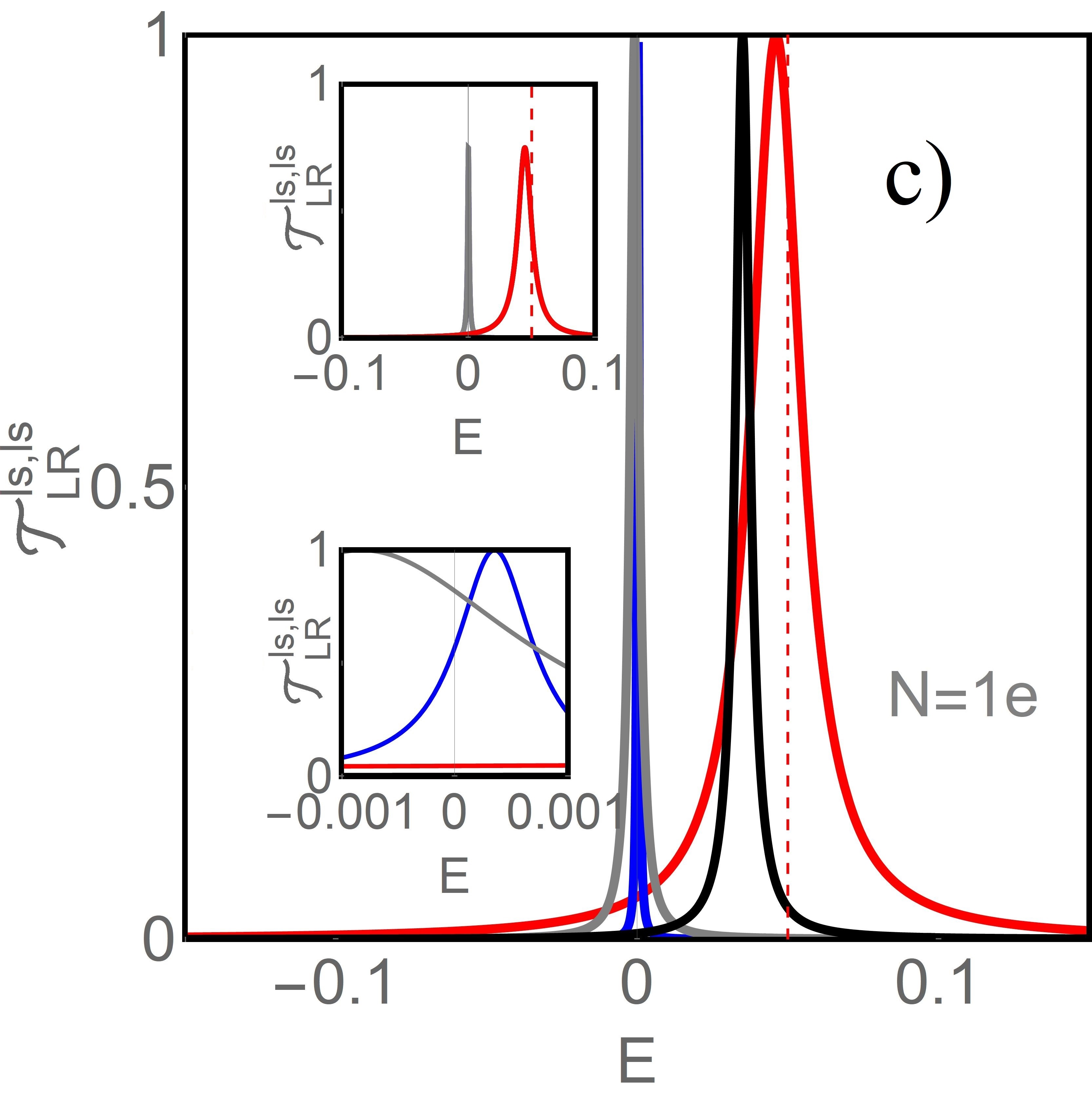}
\includegraphics[width=0.48\linewidth]{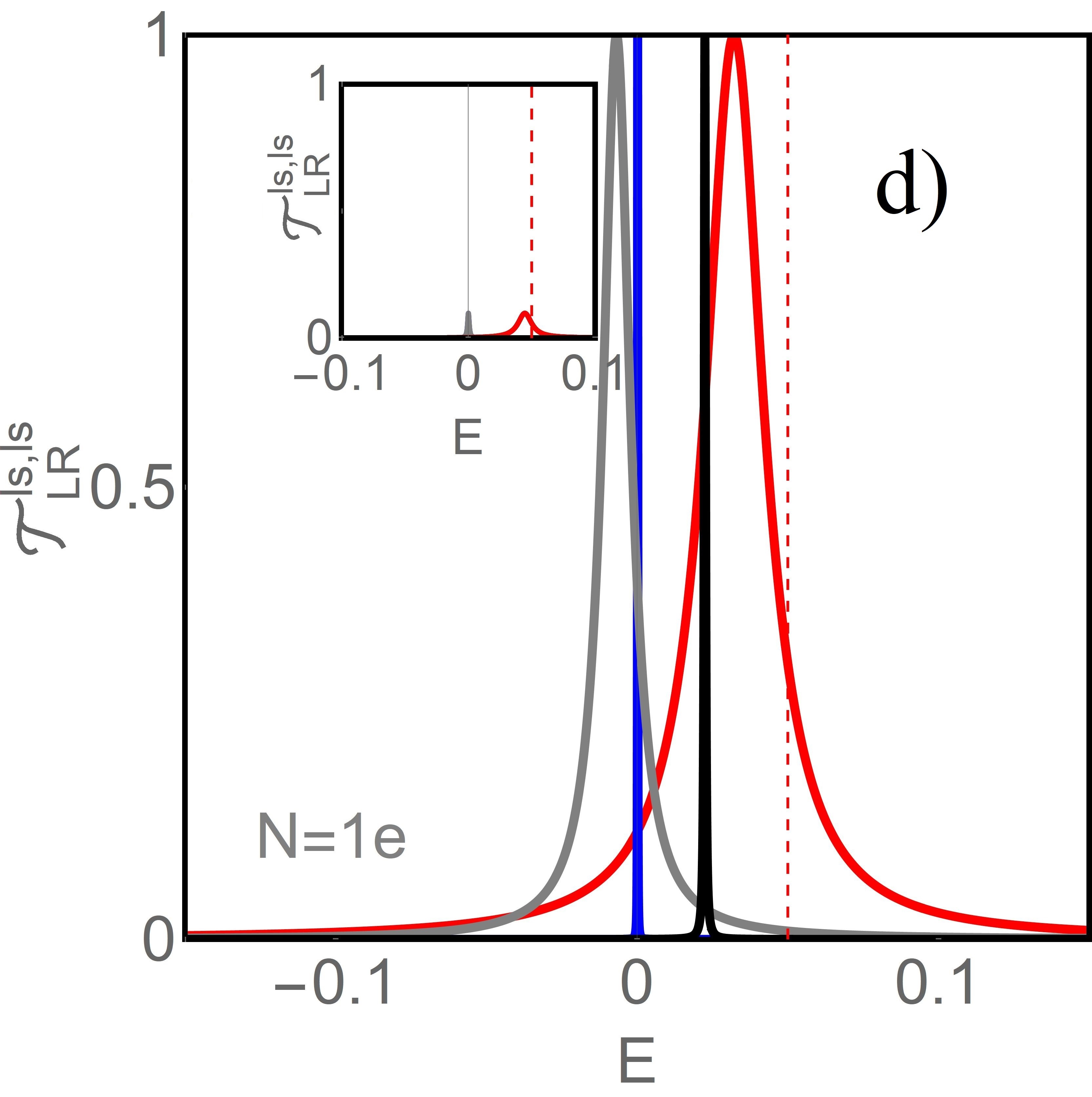}
\caption{\label{fig2} (Color online) Transmissions of CNTQD: (a,c) for $p=0.5$ and (b,d) for $p=0.95$ (red, blue black, gray lines correspond to $1\uparrow$, $1\downarrow$, $-1\uparrow$ and $-1\downarrow$ spin-orbital contributions). Upper and bottom figures are the transmissions for $\Lambda_{so}=0$ and $\Lambda_{so}=0.05$ respectively. Insets shows transmissions for antiparallel configuration. Bottom inset on Fig.2c presents the spin-orbital transmissions for P configuration close to the Fermi level (red dashed lines denote the value of the spin-orbital splitting).}
\end{figure}
where $N_{ls}=f^{\dag}_{ls}f_{ls}$ is the pseudofermion occupation operator ($d^{\dag}_{ls}\equiv z_{ls}f^{\dag}_{ls}$ \cite{Kotliar}). $Q_{ls}=p^{\dag}_{ls}p_{ls}+d^{\dag}_{l}d_{l}+d^{\dag}_{ss}d_{ss}
+d^{\dag}_{s\overline{s}}d_{s\overline{s}}+t^{\dag}_{ls}t_{ls}
+t^{\dag}_{\overline{l}s}t_{\overline{l}s}+t^{\dag}_{\overline{l}\overline{s}}t_{\overline{l}\overline{s}}
+f^{\dag}f$ and $I=e^{\dagger}e+\sum_{ls}p^{\dagger}_{ls}p_{ls}+\sum_{lss'}(d^{\dagger}_{l}d_{l}+d^{\dagger}_{ss'}d_{ss'})
+\sum_{ls}t^{\dag}_{ls}t_{ls}+f^{\dagger}f$ are the conservation and completness realtions. $z_{ls}=(e^{\dagger}p_{ls}+p^{\dagger}_{l\overline{s}}d_{l}
+p^{\dagger}_{\overline{l}\overline{s}}(\delta_{l,1}d_{s\overline{s}}+\delta_{l,-1}d_{\overline{s}s})
+p^{\dagger}_{\overline{l}s}d_{ss}+
d^{\dagger}_{\overline{l}}t_{ls}+d^{\dagger}_{\overline{s}\overline{s}}t_{\overline{l}\overline{s}}
+(\delta_{l,-1}d^{\dagger}_{s\overline{s}}+\delta_{l,1}d^{\dagger}_{\overline{s}s})t_{\overline{l}s}
+t^{\dag}_{l\overline{s}}f)/\sqrt{Q_{ls}(1-Q_{ls})}$ is the renormalization parameter
of the coupling strength with left and right electrode. Couplings to electrodes are described by matrices $\Gamma_{L}=[\Gamma^{L}_{l,s,s}]$ and $\Gamma_{R}=[\Gamma^{R}_{l,s,s'}]$, where $\Gamma^{L}_{l,s,s}= (\widetilde{\Gamma}_{ls}/2)(1\pm p)$, $\Gamma^{R}_{l,s,s}= (\widetilde{\Gamma}_{ls}/2)(1\pm p \cos(\varphi))$  for $s=\pm1$ respectively, where $\widetilde{\Gamma}_{ls}=\Gamma z^{2}_{ls}$ and $\Gamma=\pi t^{2}/2D$. The off-diagonal elements are $\Gamma^{R}_{l,s,-s}= (\widetilde{\Gamma}_{ls-s}/2)p\sin(\varphi)$ with $\widetilde{\Gamma}_{ls-s}=\Gamma z_{ls}z_{l-s}$.
$1/2D$ is the rectangular density of states in the lead ($|E|<D$ and $p$ denotes polarization of electrodes).
Throughout this paper we use relative energy units chosen $D/50$ as the unit and we set $|e| = \hbar = 1$. The spin and charge  currents are given by (${\cal{I}}^{X}=Re[{\cal{I}}^{+}]$, ${\cal{I}}^{Y}=Im[{\cal{I}}^{+}]$, ${\cal{I}}^{Z}$, ${\cal{I}}^{C}$):
\begin{eqnarray}
&&{\cal{I}}^{+}=-i\langle[\widetilde{\cal{H}},\sum_{l}ic^{\dagger}_{kLl\uparrow}c_{kLl\downarrow}-
ic^{\dagger}_{kRl\uparrow}c_{kRl\downarrow}]\rangle\\
&&\nonumber{\cal{I}}^{C(Z)}={\cal{I}}_{\uparrow}\pm {\cal{I}}_{\downarrow}=-i\langle[\widetilde{\cal{H}},\sum_{l}n_{Ll\uparrow}-n_{Rl\uparrow}
\\&&\nonumber\pm n_{Ll\downarrow}\mp n_{Rl\downarrow}]\rangle.
\end{eqnarray}
The currents are expressed by electrode-lead correlation functions, which can be found using the nonequilibrium Green functions. Finally they are given by integrals of the diagonal and off-diagonal elements of the transmission matrix ${\cal{T}}(E)=\Gamma_{L}\widetilde{G}^{R}(E)\Gamma_{R}\widetilde{G}^{A}(E)$, where $\widetilde{G}^{R(A)}$ are the retarded and advaced renornalized Green function matrix with diagonal $E-E_{ls}-\lambda_{ls}\pm i\sum_{\alpha}\Gamma^{\alpha}_{l,s,s}$ and off-diagonal elements $i\Gamma^{R}_{l,s,-s}$. Integrating the lesser Green functions, we can calculate the average spin components $S^{X,(Y,Z)}=\int \frac{\sum_{l}\widetilde{G}^{<}_{ls,ls'}(E)dE}{2\pi i}$ ($\widetilde{G}^{<}=\widetilde{G}^{R}\widetilde{\Sigma}^{<}\widetilde{G}^{A}$, where $\widetilde{\Sigma}_{ls,ls'}^{<}=2i\sum_{\alpha}\Gamma^{\alpha}_{l,s,s'}f_{\alpha}(E)$, $f_{\alpha}(E)$ is the Fermi distribution function)\cite{Barnas}. The differential conductance, tunnel magneto-resistance (TMR) and polarization of conductance are given by formulas: $TMR(\varphi)=[{\cal{G}}(0)-{\cal{G}}(\varphi)]/{\cal{G}}(\varphi)$ and $PC=[{\cal{G}}_{\uparrow}-{\cal{G}}_{\downarrow}]/{\cal{G}}$ (where ${\cal{G}}=\sum_{s}{\cal{G}}_{s}=\sum_{s}d{\cal{I}}_{s}/dV=(e^{2}/h)\int^{+\infty}_{-\infty}(f_{L}-f_{R})\sum_{ls}{\cal{T}}^{ls,ls}_{LR}(E)dE$. All the calculations were performed for $\Gamma=0.05$ and $U=6$.

\section{Results and discussion}
The major issue of spintronics is a control of transport by a change of relative orientations of magnetic moments of external leads. Fig. 1b shows the gate voltage dependence of tunnel magnetoresistance of CNT-QD and Figures 1d,e present the corresponding conductances for P and AP configurations.
In collinear configuration of magnetizations of the leads the decisive role in linear TMR plays a competition of the central transmission peak (peak  located closest to the Fermi level) of the diagonal transmission for parallel (P) configuration (Fig. 2a,b) and central peak for antiparallel (AP) orientation (insets of Fig. 2a,b). The majority peak is wider and located closer to the Fermi level. In AP configuration ($\varphi = \pi$) only a single peak is visible with equal contributions of both spin orientations. As it is seen from comparison of Figs. 2a,b the disproportion between P and AP transmissions increases with increasing electrode polarization P, which means an increase in TMR. For $p = 1$  AP transmission at the Fermi level disappears and transport is blocked (perfect spin valve effect). Without SO coupling ($\Lambda_{so} = 0$) TMR reaches value only  weakly dependent on gate voltage and determined by polarization of electrodes $\frac{p^{2}}{1-p^{2}}$ \cite{Moca} and it increases up to Julli\`{e}re limit $\frac{2p^{2}}{1-p^{2}}$ when approaches  fully occupied or empty  regions of the dot.
For clarification of gate voltage dependencies of TMR  we present on Figs.1d, ecorresponding conductances for P and AP configurations. For $\Lambda_{so}=0$   AP conductance curve resembles dependence for $p=0$, but with  reduced plateau values for all occupations.  This is a consequence of breaking of the  left-right symmetry in individual spin channels. For P orientation differences between  spin-resolved conductances  result from violation of  spin symmetry. For $n=2$ exchange splitting $\Delta$ is small and vanishes at e-h symmetry point (inset of Fig. 3b). Transmissions  for both spins  locate close to $E_{F}$ and both  spin resolved conductances reach almost unitary limits in the centre of $n=2$ valley. Outside this point they behave differently due to the different widths of transmissions.  For $n=1,3$ transmission lines  do not locate at $E_{F}$ and therefore the points of vanishing $\Delta$ do not reflect in equality of  conductances. For $\Lambda_{so}\neq0$  in  P configuration conductances differ for different spin and orbital channels. Breaking of SU(4) symmetry  reflects most strongly by suppression of conductances for $n=2$, because in this region SU(4) Kondo temperature  ($p=0$, $\Lambda_{so}=0$) is lowest.  In AP configuration there are only two different orbital resolved conductances, which are mirror reflections with respect to the e-h symmetry line. Independent on the sign of spin splitting $\Delta=E_{l+}+\lambda_{l+}-E_{l-}-\lambda_{l-}$ conductance for parallel orientation of the  magnetic moments of electrodes dominates over conductance for antiparallel orientation and   TMR remains positive in the whole range of gate potential. This changes for non-zero SO coupling.
\begin{figure}[h!]
\includegraphics[width=0.48\linewidth]{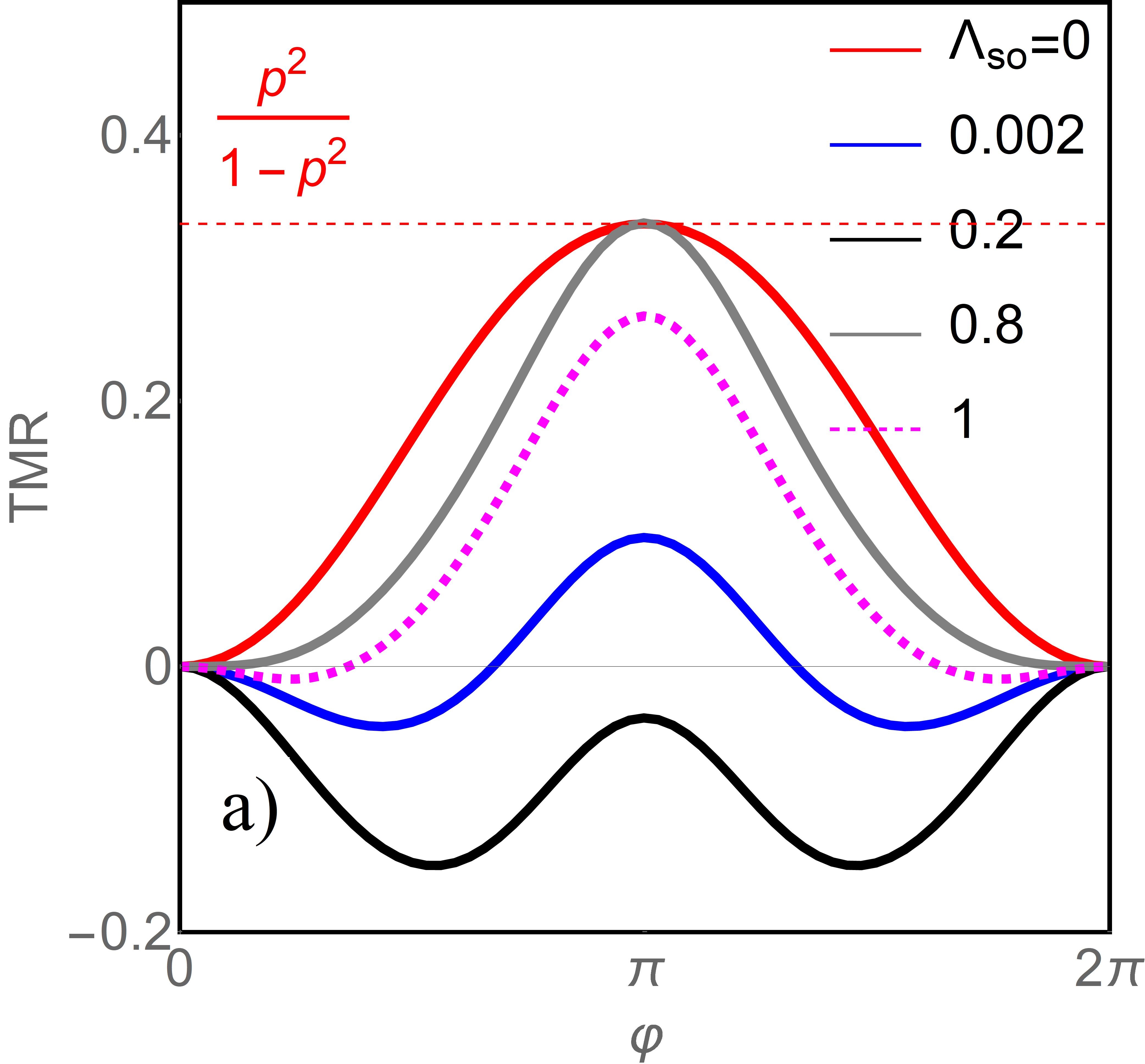}
\includegraphics[width=0.48\linewidth]{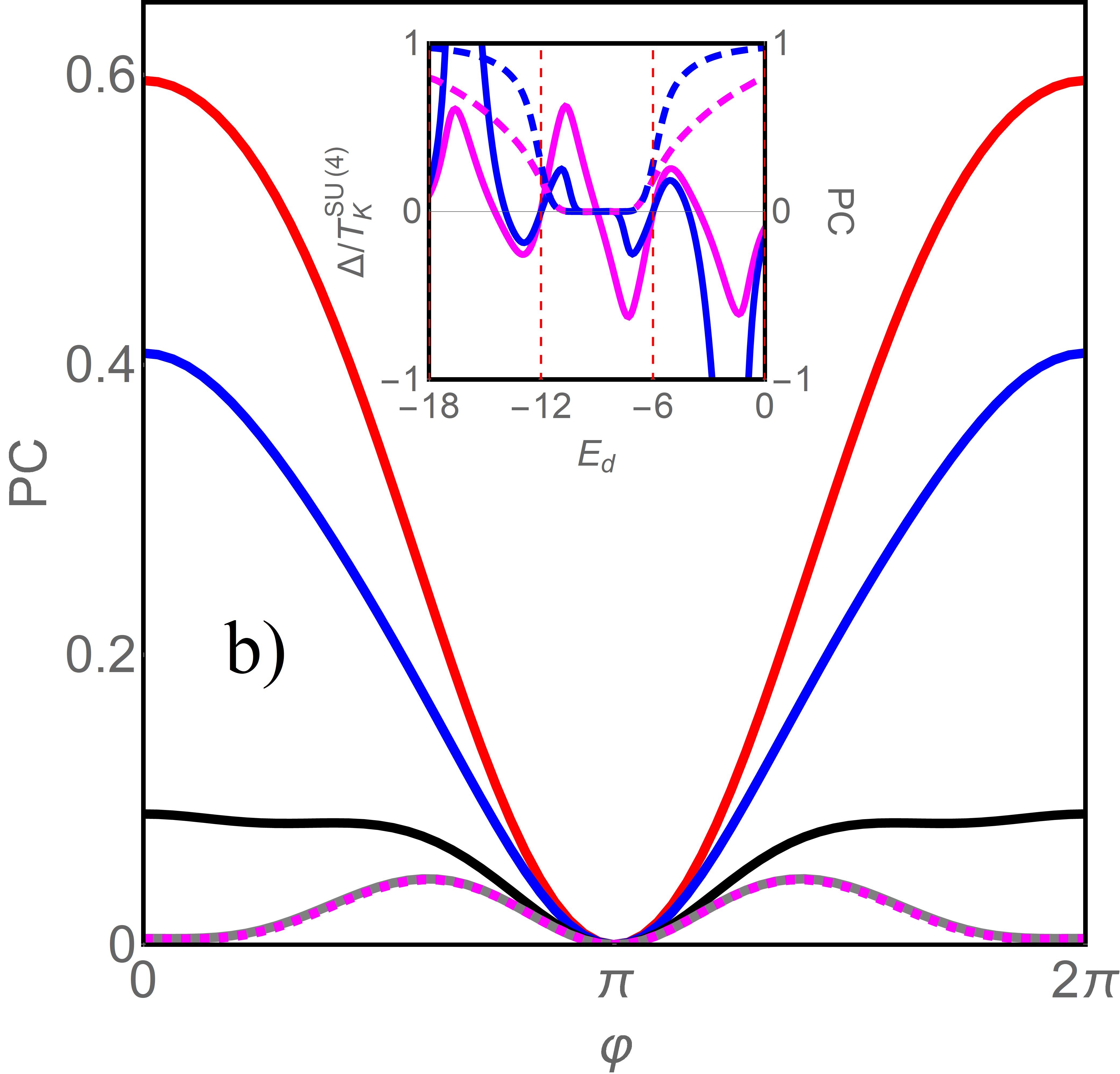}\\
\includegraphics[width=0.48\linewidth]{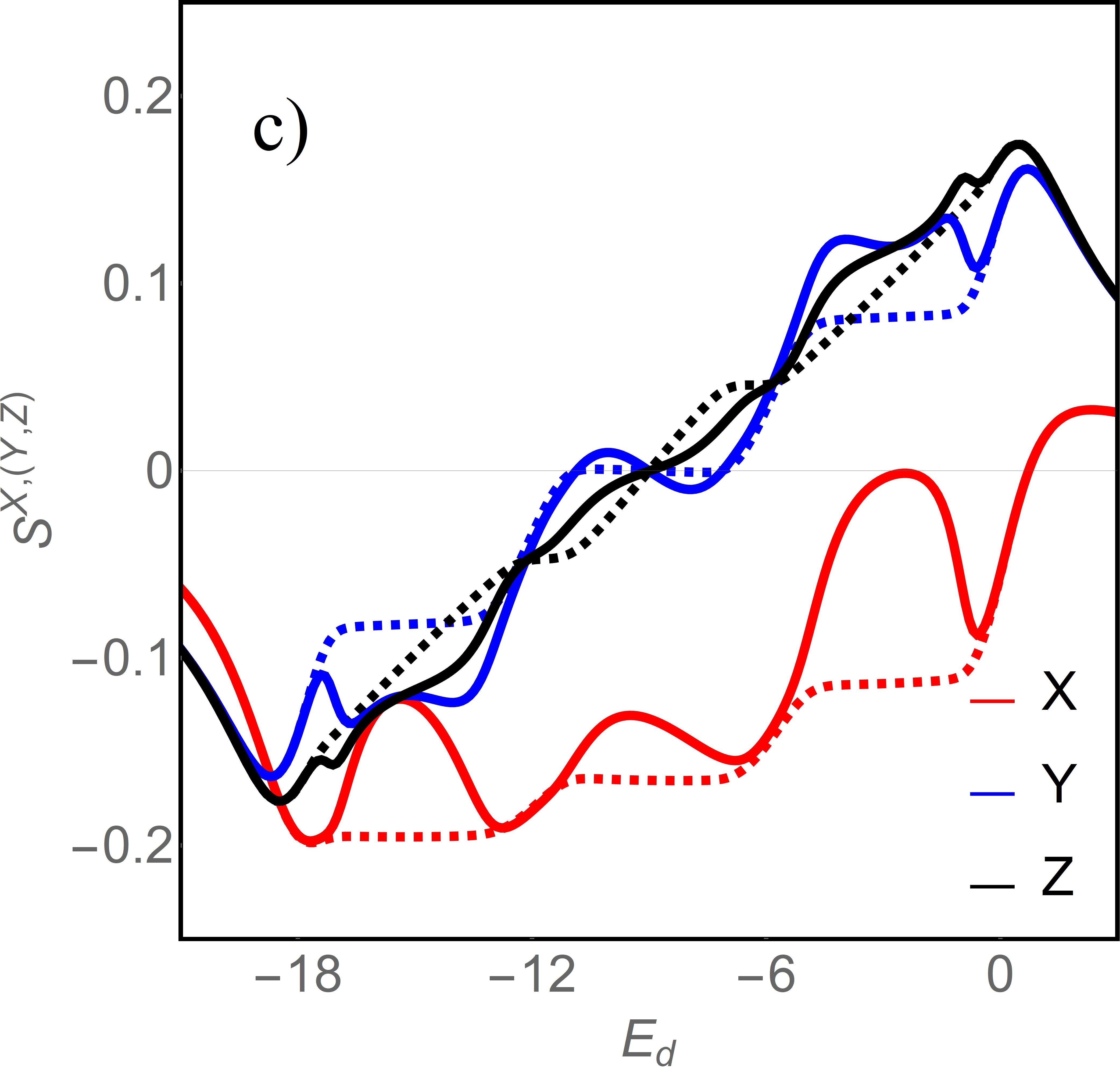}
\includegraphics[width=0.48\linewidth]{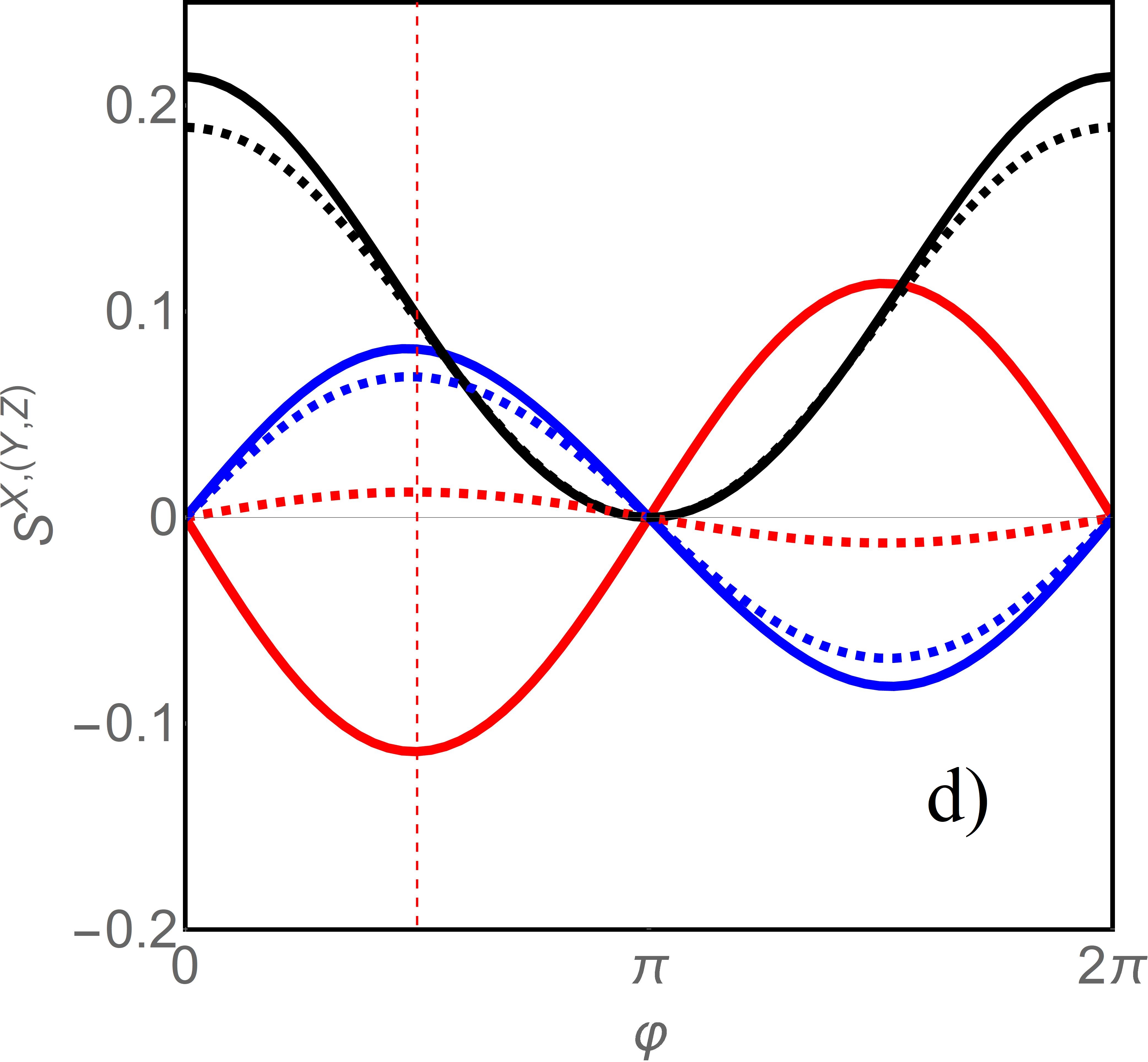}\\
\includegraphics[width=0.48\linewidth]{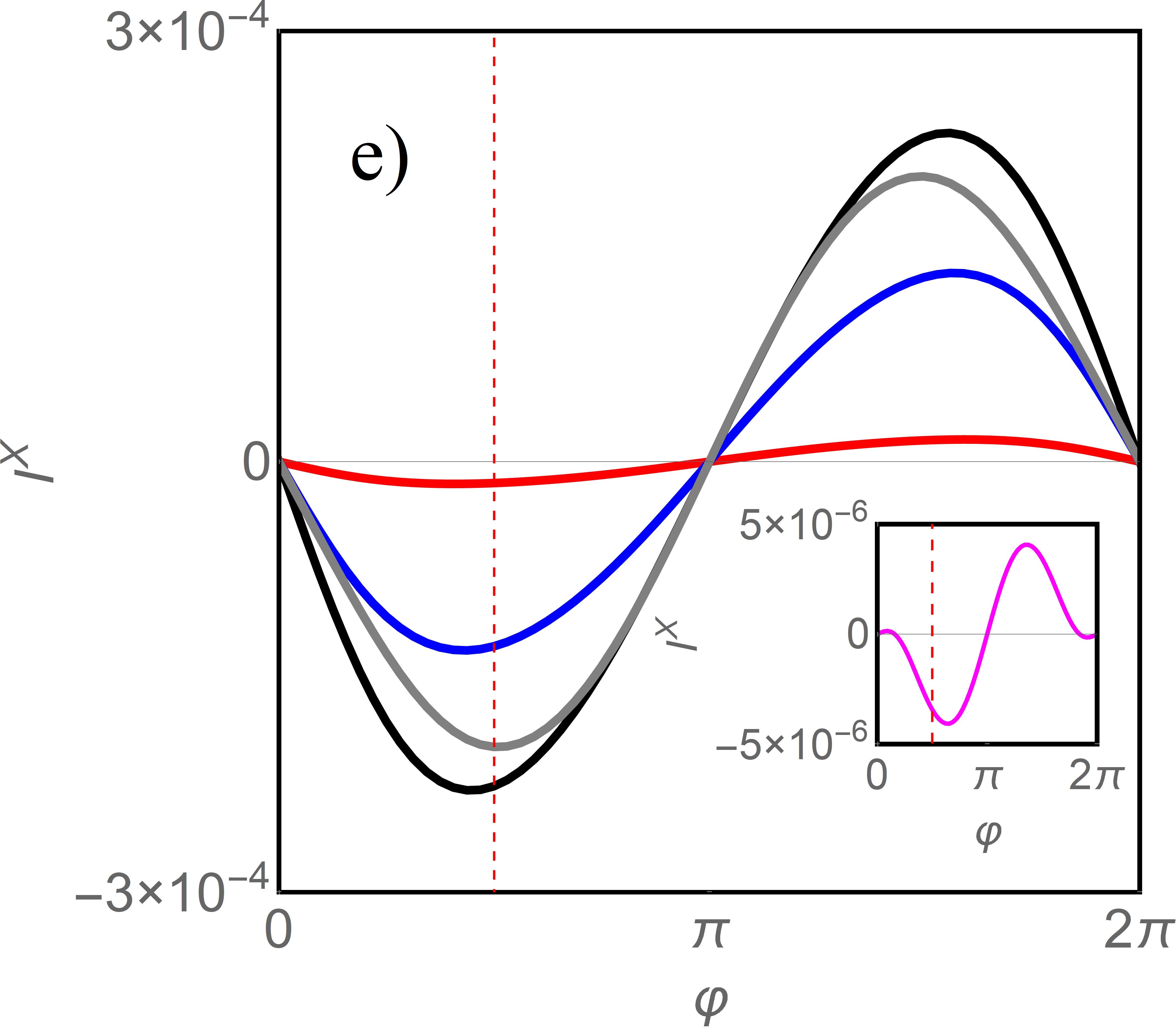}
\caption{\label{fig3} (Color online) (a,b) TMR and PC as a function of angle $\varphi$ presented for different SOC strengths. Inset of Fig.b shows gate-dependent exchange field splitting ($\Delta/T^{SU(4)}_{K}$) and PC for $p=0.5,0.8$ (solid and dashed, magenta and blue lines).(c) - gate dependencies of average spin components for $\Lambda_{so}=0$ (dotted lines) and for $\Lambda_{so}=0.05$ (solid lines) ($\varphi=\pi/2$). d) $\langle S^{X,(Y,Z)}\rangle$ as a function of $\varphi$ for $\Lambda_{so}=0$ (solid lines) and for $\Lambda_{so}=0.3$ (dotted lines). (e) - Angle dependencies of transverse equilibrium spin currents for $\Lambda_{so}=0, 0.01, 0.05, 0.3$ (red, blue, black and gray lines). Inset illustrates ${\cal{I}}^{X}$ for $\Lambda_{so}=1$. Dashed lines mark $\varphi=\pi/2$ ($p=0.5$).}
\end{figure}
In transmission for P configuration in this case (Fig. 2c,d), additional satellite peak appears located approximately around $E = \Delta\pm\Lambda_{so}$ in $N = 1e$ region or $E = -\Delta\mp\Lambda_{so}$ for $N = 3e$  and $E=\pm \Lambda_{so}$ for $N = 2e$. In the upper inset of Fig. 2d presenting transmission for AP orientation only satellite corresponding to the SO splitting is visible. The reconstruction of transmissions in P and AP configurations causes  a suppression of TMR and for $\Lambda_{so}\neq0$  even the sign of magnetoresistance may change in some cases (inverse TMR). Fig. 1c illustrates what a strong effect on TMR has weakening of the coupling of the leads to the dot.
As it is seen with the decrease of $\Gamma$ inverse TMR($\varphi$) appears at lower value of SOC.
This is a consequence of narrowing of perturbed  Kondo peaks and in consequence the   spin-orbit induced splitting of transmission peaks  occurs for smaller values of $\Lambda_{so}$. Fig. 3a   presents the angle dependence of TMR. When the directions of electrode magnetizations deviate from each other, TMR increases and reaches maximal value for $\varphi = \pi$. SO coupling decreases the amplitude of TMR oscillation and, as we already mentioned earlier, allows for its negative values. Inverse TMR is observed already for very small values of SOC ($\Lambda_{so}=0.002$) and maximal value reaches for $\Lambda_{so}=0.2$ for noncollinear configuration ($\varphi=\pi/2$ and $\varphi=3\pi/2$). Fig. 3b shows the oscillating, decreasing with the increase of  $\Lambda_{so}$,  angle dependence of polarization of current with maximum for P configuration, what is a direct consequence of highest transmission for spin up channel for this configuration  (Fig.2). Despite the fact that exchange field changes sign when moving form $N = 1e$ region  into $N = 3e$  PC remains positive for all gate voltages (inset of Fig. 3b).
Kondo temperature $T^{SU(4)}_{K}$ changes with gate voltage and it ranges from $10^{-5}$ to $10^{-3}$.
\begin{figure}[h!]
\includegraphics[width=0.48\linewidth]{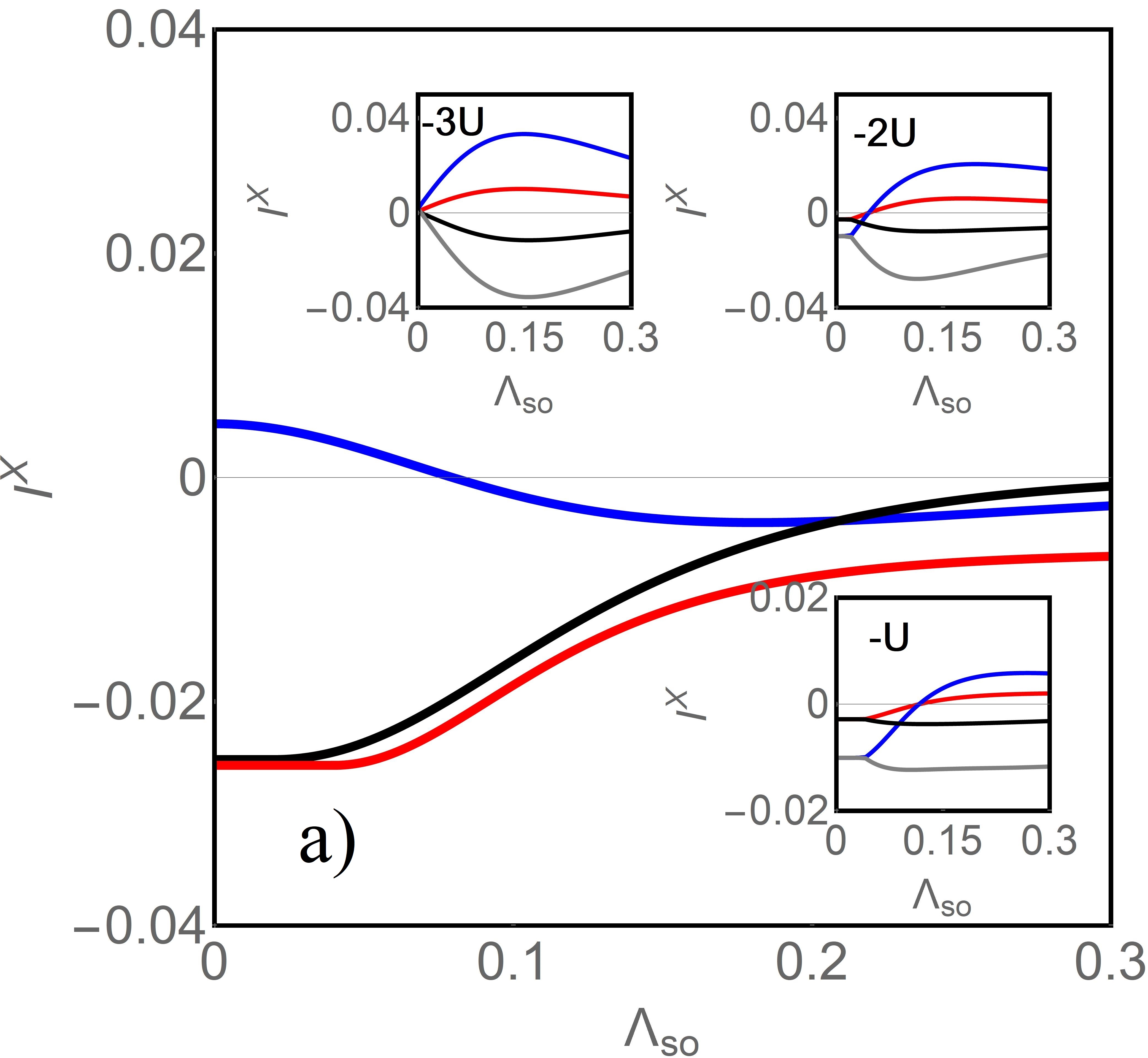}
\includegraphics[width=0.48\linewidth]{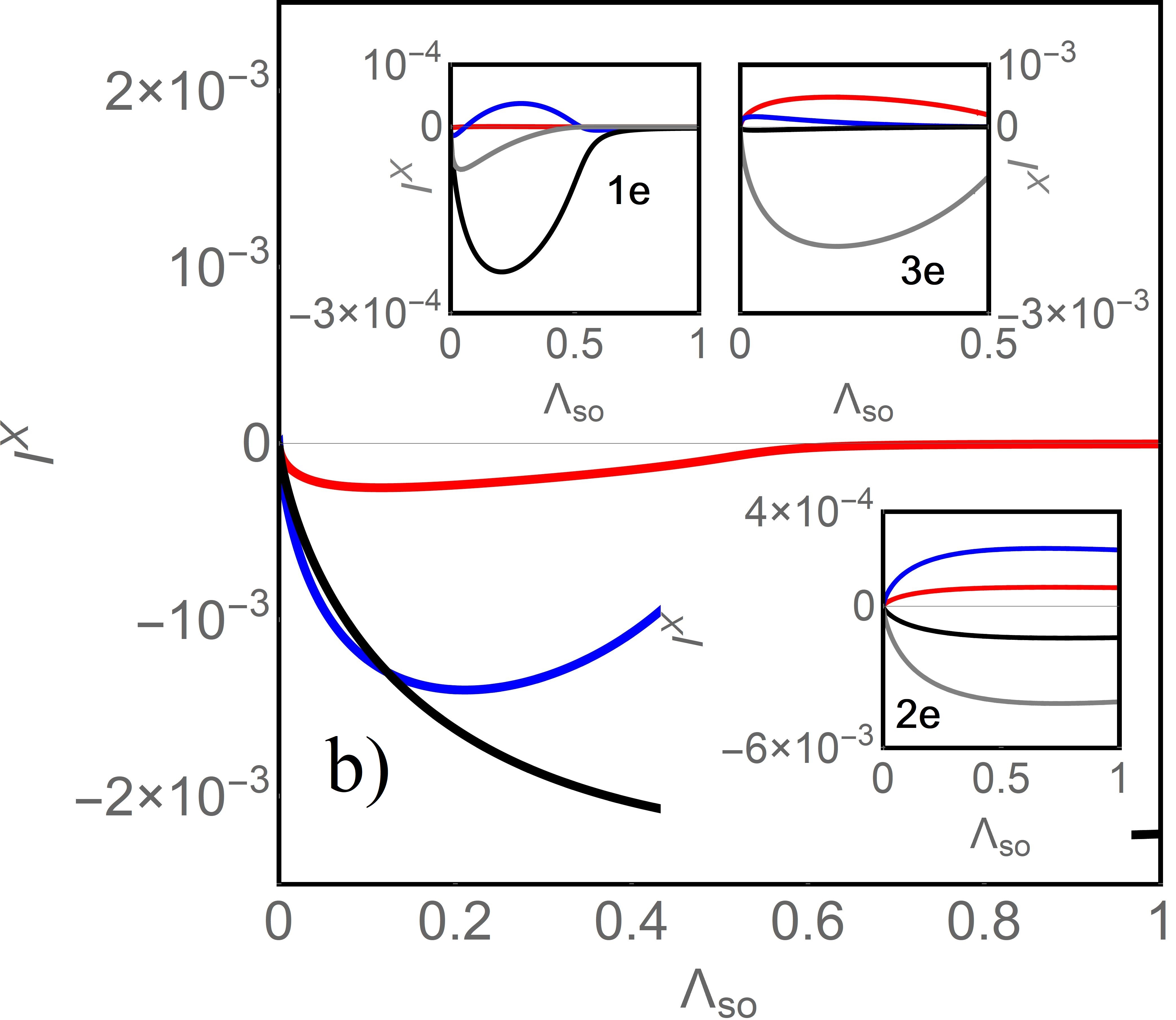}\\
\includegraphics[width=0.48\linewidth]{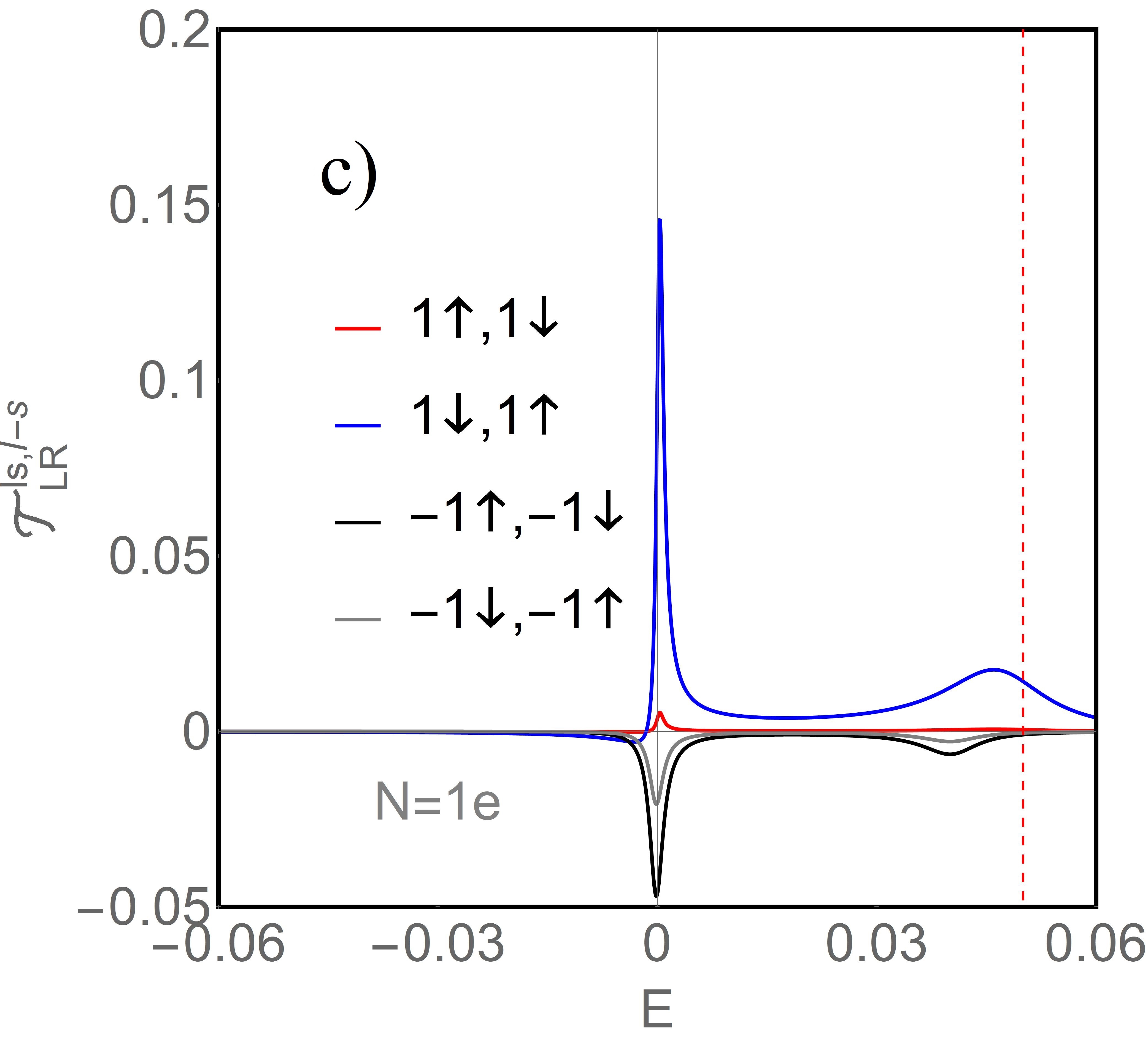}
\includegraphics[width=0.48\linewidth]{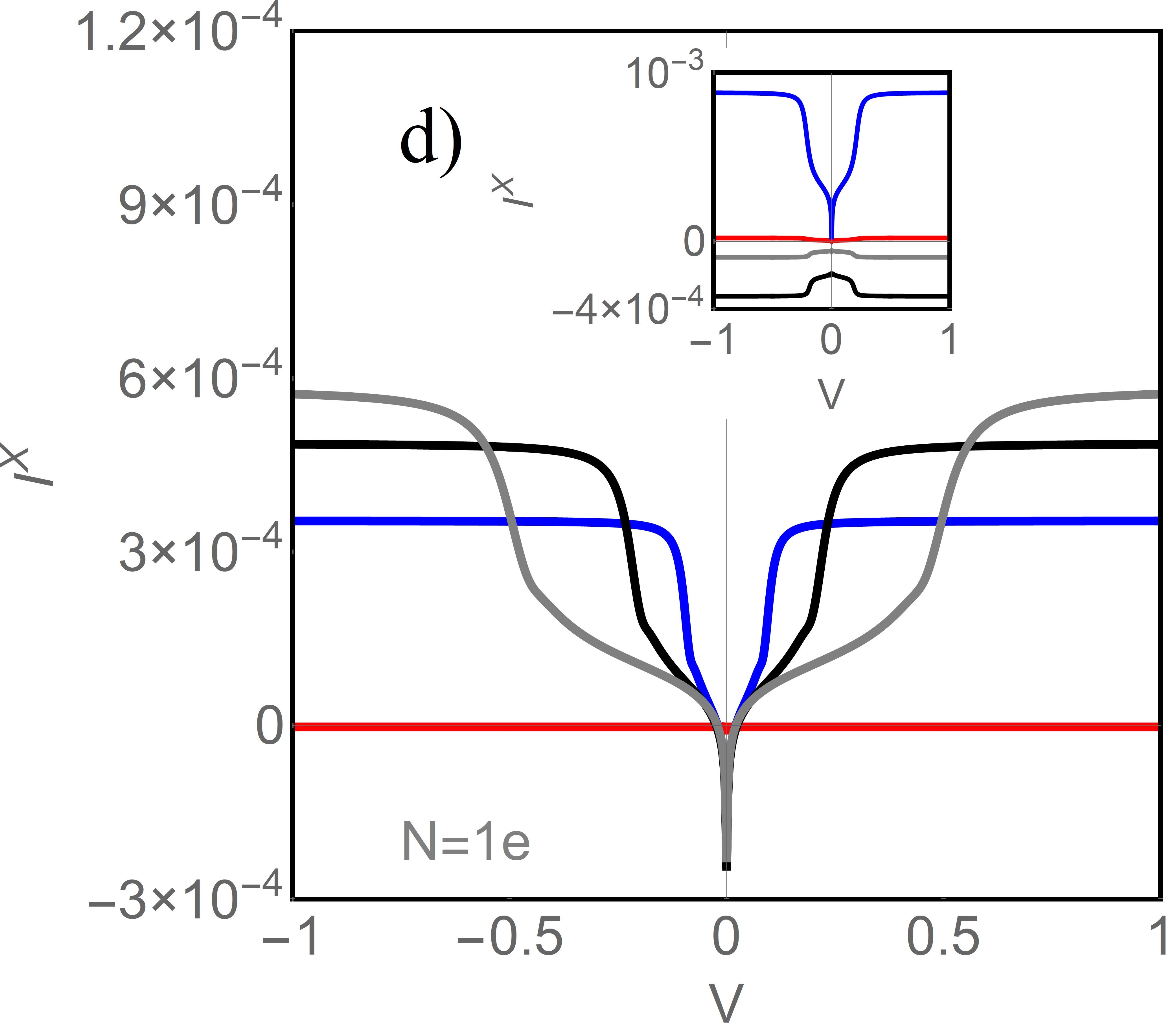}\\
\includegraphics[width=0.48\linewidth]{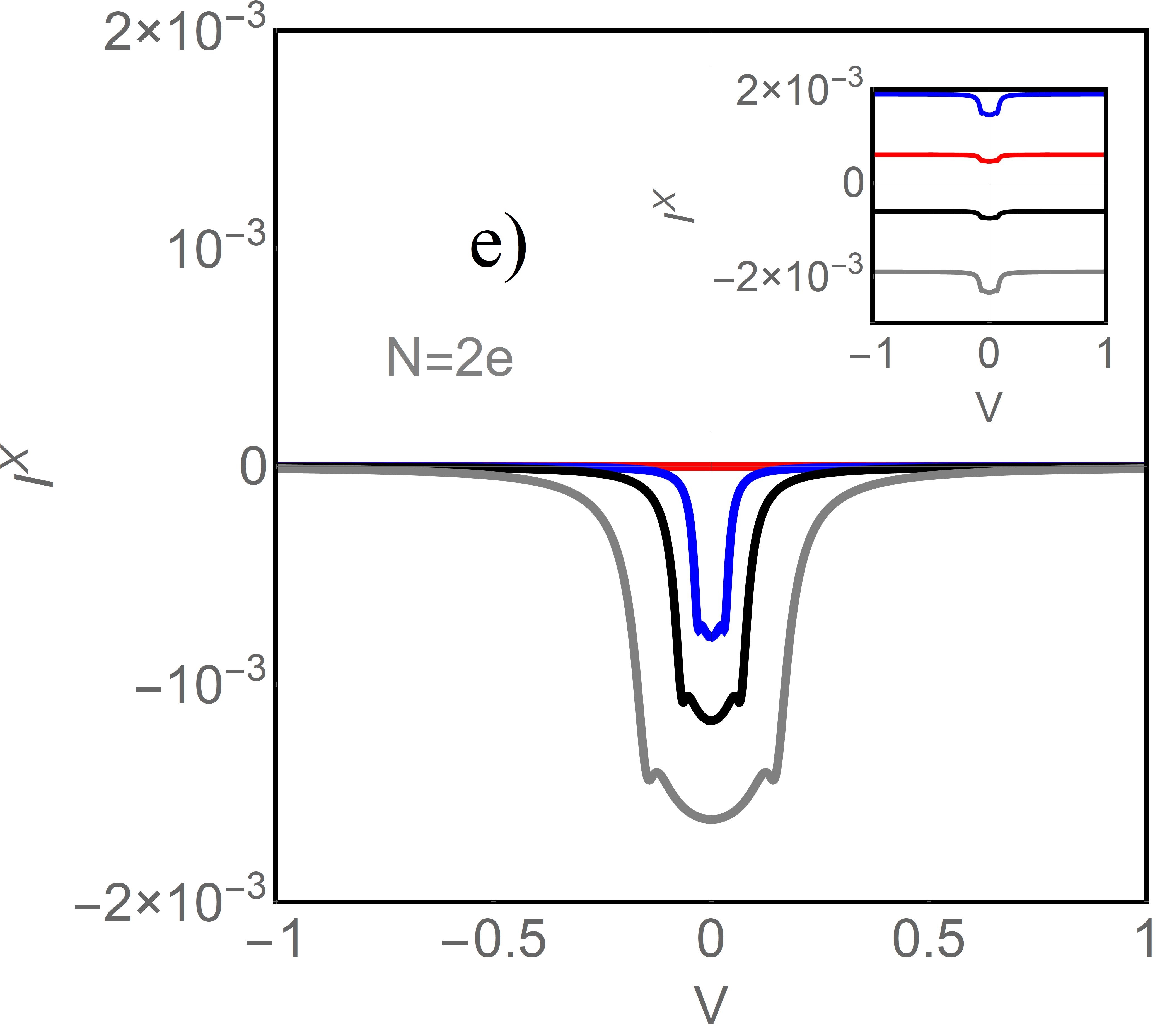}
\includegraphics[width=0.48\linewidth]{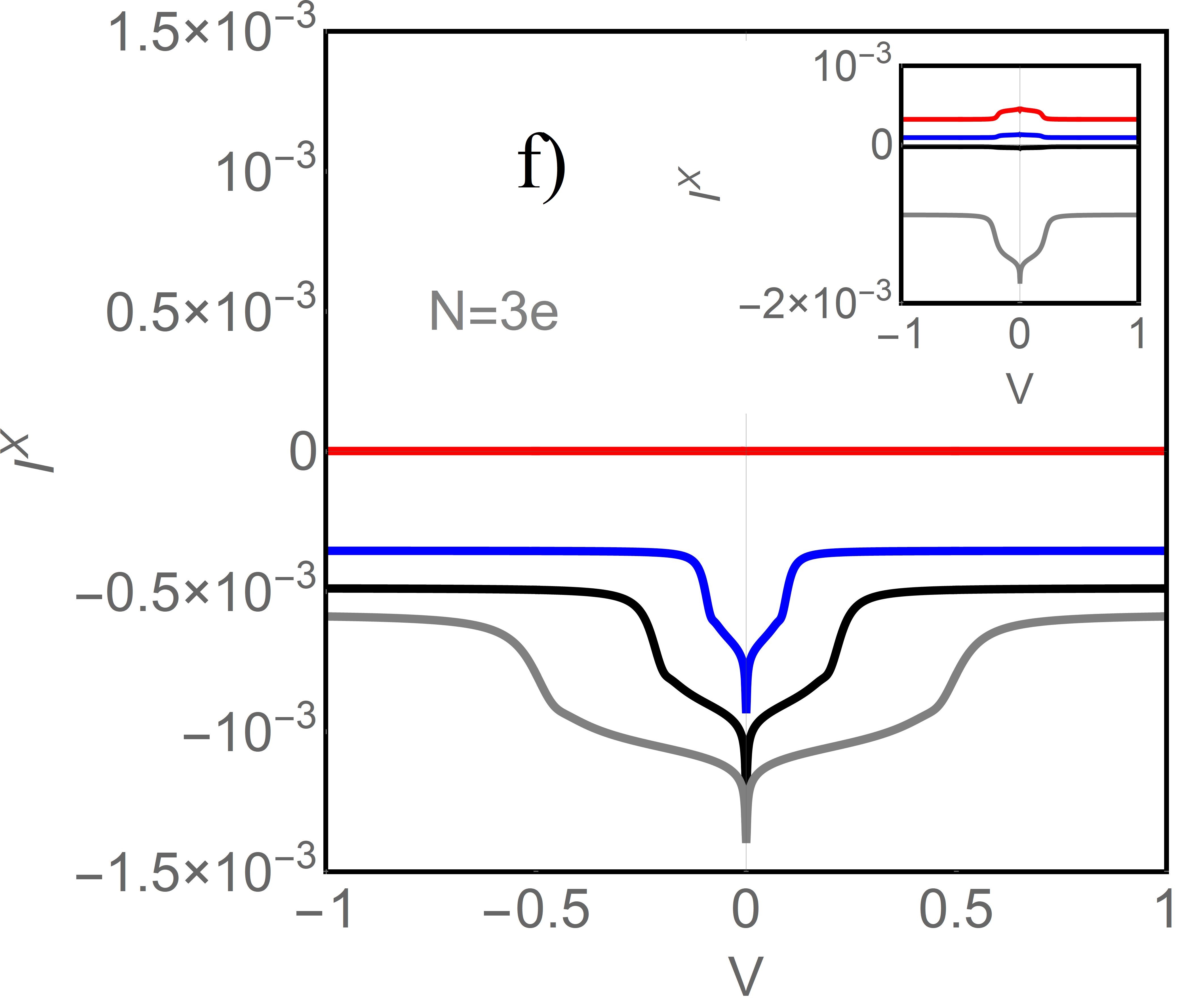}
\caption{\label{fig4} (Color online) (a) Equlibrium transverse spin currents ${\cal{I}}^{X}$ at the borders of Coulomb blockades ($E_{d}=-U,-2U,-3U$ - red, black, blue curves). (b) ESC in Kondo range for $N=1,2,3e$ (red, black, blue lines). Insets present orbital spin-flip  contributions to equilibrium spin currents (${\cal{I}}^{1\uparrow,1\downarrow}$, ${\cal{I}}^{1\downarrow,1\uparrow}$, ${\cal{I}}^{-1\uparrow,-1\downarrow}$ and ${\cal{I}}^{-1\downarrow,-1\uparrow}$ - red, blue, black, gray curves). c) The off-diagonal spin-flip orbital transmissions for $N=1e$. Red dashed line marks $\Lambda_{so}=0.05$.  d-f) ${\cal{I}}^{X}$ as a function of bias voltage for $N=1e,2e,3e$ shown for $\Lambda_{so}=0,0.05,0.1,0.2$ (red, blue, black and gray lines). Insets present orbital spin-flip contributions to spin currents for $\Lambda=0.1$ ($\varphi=\pi/2$, $p=1/2$).}
\end{figure}
Coupling of CNTQD to polarized electrodes induces  spin accumulation. The average values of  spin components plotted as a functions of gate voltage for $\Lambda_{so} = 0$ and for nonvanishing SO coupling are shown on Fig. 3c. Due to exchange field induced by polarized electrodes the dot magnetic moment is suppressed, but not   quenched and local extrema occur at the boundaries of Coulomb blockades. The change of  the sign  of transverse spin components when crossing electron-hole symmetry point is the consequence of reversal  of the  effective exchange field.
The angle  dependencies of average spin components are shown on Fig. 3d. $\langle S^{Z}\rangle$ vanishes for antiparallel configuration ($\varphi = \pi$), whereas transverse components $\langle S^{X,Y}\rangle$ vanish for $\varphi = 0$ and $\varphi = \pi$. SO coupling diminishes the oscillation amplitude. Recently there has been an increasing interest  in generation of pure spin currents without an accompanying charge current.  Fig. 3e presents the angle dependencies of equilibrium (zero bias) transverse component of spin current ${\cal{I}}^{X}$ (ESC). When magnetic moments of electrodes are deviating from the parallel orientation the transverse components of the exchange field appear. For the geometry considered it is $Y$ component of the effective field ($H^{Y}$) and spin torque  $H^{Y}\times s$ pointing in $X$ direction exerts on electrons flowing through the dot ($s = \uparrow(\downarrow)$ in the global Z axis). Torque is oriented in the opposite directions for right and left moving electrons. In consequence, the charge flow in opposite directions is associated with opposite x components of the spin. Fig. 4 concerns AP configuration. Fig. 4a  presents dependencies of ESC  on SO coupling strength at the borders of Coulomb blockades and Fig. 4b the same dependencies for  the regions of broken Kondo states for occupations $N =1,2,3$. In the former case absolute values of  ${\cal{I}}^{X}$  decrease with the increase of SO coupling, whereas for the latter nonmonotonic dependencies are observed with predominantly increasing tendencies of ${\cal{I}}^{X}$ for small values  of $\Lambda_{so}$. Although the behavior of ESC is determined by a subtle interplay of integrated  to the Fermi level off-diagonal in spin space transmissions, the increasing tendency in Kondo regime reflects the general property of increase of currents due to the increase of electrode-dot coupling with SO induced weakening of Kondo correlations. Fig. 4c shows examples of  zero-bias partial off-diagonal transmissions for $N = 1$. $T_{LR}^{-1\uparrow,-1\downarrow}$ dominates at $E_{F}$ and so does the corresponding integral up to $E_{F}$ determining  spin-flip ${\cal{I}}^{-1\uparrow,-1\downarrow}$  contribution to ${\cal{I}}^{X}$. Its largest and negative contribution is shown in the left upper inset of Fig. 4b. Similar partial, orbital resolved spin-flip equilibrium spin currents for selected  gate voltages are presented on the rest of insets of Figs. 4a,b.  Figs. 4 d,e,f  present bias voltage dependencies of ${\cal{I}}^{X}$  spin current for different occupation regimes. Independent on the value of $\Lambda_{so}$, the spin current remains negative for $N = 2$ and $N = 3$, whereas for $N = 1$ it changes sign at small voltages. For  the detailed explanation of this observation  the picture of the evolution of spin-flip transmissions with voltage should be recalled, but even from the insight into the  zero-bias transmissions (Fig. 4 and its inset) it is seen  that  by increasing the transport window the positive contribution ${\cal{I}}^{1\downarrow,1\uparrow}$ starts to dominate in ${\cal{I}}^{X}$ over the negative.

Summarizing, we have investigated the effect of spin-orbit coupling and Kondo correlations on spin currents and tunnel magnetoresistance in carbon nanotube quantum dot comparing also some calculations with the results for the boundaries between the Coulomb valleys. Discussion of SO effects and noncollinearity of polarizations is important because the energy of SO coupling is comparable with the energy scale of Kondo effect and deviations of polarizations introduce real spin-flip processes. In consequence  the many body resonances  are formed due to the interplay between Kondo-like effective spin-orbital flips and real spin-orbit transitions.   Apart from the current with spin component parallel to the global quantization axis, noncollinearity of polarizations induces also transverse components (spin-flip currents), which do not vanish even in equilibrium.  ESC comes from the exchange coupling  between the magnetic moments of ferromagnetic electrodes and its direction is determined by the vector product of these magnetizations.   The spin currents scale with coupling strengths to the leads and as such are much weaker in the Kondo range than in Coulomb charge fluctuation regions. Spin-orbit coupling weakens TMR, but  due to this interaction the inverse TMR effect may occur.

\def\refname{References}

\end{document}